\documentclass[runningheads,a4paper]{llncs}
\usepackage[T1]{fontenc}
\usepackage[utf8]{inputenc}

\usepackage[lighttt]{lmodern}
\usepackage{bm}

\usepackage{gstyle}
\usepackage{gtt}

\usepackage{blindtext}

\usepackage{aofvauditing}
\author{
  Alexander Weigl\inst{1}
  \and Mattias Ulbrich\inst{1}
  \and Suhyun Cha\inst{2}
  \and Bernhard Beckert\inst{1}
  \and Birgit Vogel-Heuser\inst{2}
}

\authorrunning{
  A. Weigl et.\,al.
}

\institute{Karlsruhe Institute of Technology, Germany \\
  \email{beckert@kit.edu}, \email{ulbrich@kit.edu}, \email{weigl@kit.edu}
  \and
  Technical University of Munich, Germany\\
  \email{suhyun.cha@tum.de}, \email{vogel-heuser@ais.mw.tum.de}
}

\title{Relational Test Tables: A Practical Specification Language for Evolution
  and Security\thanks{Research supported by the DFG (German Research Foundation)
    in Priority Programme SPP1593: Design for Future -- Managed Software
    Evolution (VO~937/28-2, BE~2334/7-2, and UL~433/1-2).}}

\titlerunning{Relational Test Tables}%

\renewcommand\paragraph[1]{\myparagraph{#1}}

\newcommand\IF{information flow\xspace}

\begin{document}
\pagestyle{headings}
\mainmatter

\maketitle

\begin{abstract} %
  A wide range of interesting program properties are intrinsically relational,
  i.e., they relate two or more program traces. Two prominent relational
  properties are \emph{secure \IF} and \emph{conditional program equivalence}.
  By showing the absence of illegal \IF, confidentiality and
  integrity properties can be proved. Equivalence proofs allow using
  an existing (trusted) software release as specification for new revisions.

  Currently, the verification of relational properties is hardly accessible to
  practitioners due to the lack of appropriate relational specification
  languages.

  In previous work, we introduced the concept of generalised test tables:
  a table-based specification language for \emph{functional} (non-relational)
  properties of reactive
  systems. In this paper, we present \emph{relational} test tables --
  a canonical extension of generalised test tables for the specification of
  relational properties, which refer to two or more program runs or
  traces. Regression test tables support asynchronous program runs via
  stuttering. We show the applicability of relational test tables, using them
  for the specification and verification of two examples from the domain of
  automated product systems.
\end{abstract}

\section{Introduction}
\label{sec:introduction}

\myparagraph{Motivation}
Relational specifications allow formalising interesting and practically
relevant properties.
Two important applications of proving relational program properties are
regression verification and ensuring secure \IF or non-interference.
Regression verification is a generalisation of program equivalence proofs, where
two program revisions are shown to be related without requiring full
equivalence~\cite{BeckertUlbrichEtAl2015}.
Secure \IF and non-interference require that a specified (secret) input values
do not have an effect on some (public) output variables~\cite{Denning}.
Both properties -- equivalence and information flow -- 
are specified as a relation between program runs.
For example, to describe the equivalence of two versions of a reactive system, 
we may state that traces of the two versions are 
state-wise equivalent in their output values
if they are equivalent in their input values.
In general, a relational property~$r$ for $n\geq 2$~reactive systems is
formalised as a universal quantification
$\forall t_1\ldots\forall t_n\,.\, \mathit{r}(t_1,\ldots, t_n)$, where $t_i$ is
an infinite trace of the $i$th system~\cite{Clarkson2010}.

Currently, formal methods for proving relational properties of reactive
systems are rarely used in practice. One of the main obstacles is the lack of
appropriated specification languages (the situation is similar as with
non-relational one-trace properties~\cite{PaVy:16}).

In~\cite{Beckert2017}, we introduced the concept of generalised test tables
(\gtts): a table-based specification language for functional specification of
reactive systems, which combines the comprehensibility of concrete test tables
with the expressiveness of formal specifications.
In contrast to concrete test tables (which are used in industry), each \gtt{}
specifies a whole family or class of test cases.
The advantage of \gtts is the broken down view on variables and time. In
comparison to LTL, a table makes it very clear to the user, which variables
a given constraint restricts and in which order constraints become relevant
during program execution.

The observable behaviour of a reactive system is not only a pair of
input/output values but an entire trace (unlike, e.g., for algorithmic
computations). Hence, relational specifications cannot only set values
of the same step into relation, but can also span over multiple steps,
e.g., by allowing one system to \emph{stutter} (staying in the same
state for several cycles while the other systems evolve). It may be a
sensible thing to specify, e.g., that one system stays in one state as
least as long as the other system before they continue synchronously.
In contrast to Bisimulation and the simulation hierarchy, a relational test
table is defined upon traces and not a Kripke structure.
There might be some intersection simulation hierarchy and specification, which
are expressible in relational test tables, but in general these domains are
disjoint.

The presented relational specification language can be applied versatilely: for
formal verification (e.g., using a model checker) as presented in
Sect.~\ref{sec:applicaton}, for test case generation or for run time monitoring.

\myparagraph{Contribution} In this paper, we propose a canonical extension of
\gtts: \emph{relational} test tables (\rtts).
\Rtts allow the specification of relational (and functional)
properties of reactive systems in a practical and comprehensive
formalism. They can be used for relations on any number $n\geq 2$ of
system traces.  We present the syntax (Sect.~\ref{sec:syntax}) and the
semantics (Sect.~\ref{sec:semantics}) of the extension, including the
\emph{stuttering}.
Moreover, we show the applicability of \rtts with two examples
from the domain of automated product systems; one example demonstrates the use
of \rtts for regression verification and one the use for specifying
non-interference (Sect.~\ref{sec:applicaton}). Both examples have been 
successfully verified using a model checker.

\paragraph{Related Work}
\label{sec:related-work}
Clarkson et.\,al.\ propose extension of LTL and CTL$^*$ for the specification of
hyperproperties on Kripke structures~\cite{DBLP:conf/post/ClarksonFKMRS14}.
Both temporal logics are extended by existential and universal trace quantifier,
and proposition are denoted to a particular trace (cf.~Sect.~\ref{sec:syntax}).
In comparison, due to the existential quantification, HyperLTL and HyperCTL$^*$
allows the specification of a super set of relational properties.
On the other side, they inherit the problems from LTL and CTL$^*$ for the
practical use.
Barthe et. al. \cite{DBLP:journals/corr/abs-1906-09899} uses first-order logic
(FOL) to express hyper properties, e.\,g. non-interference. The FOL signature
contains the theory of natural numbers and integer, and also includes symbols
denoting timepoints, last iterations in loops, program variables and traces.

There are several extensions for specifications appraoches in the deductive
verification domain.
Yi et.\,al. propose change contracts in~\cite{DBLP:journals/tosem/YiQTR15}.
Change contracts are an extension to the Java modeling language (JML) which
allows the description of the behavioural as well as structural changes between
two methods.
Similarly, Scheben and Schmitt~\cite{DBLP:conf/fm/SchebenS14} presents a JML
extension for the specification of secure \IF.
Blatter\,et.\,al.\, \cite{Blatter2016DeductiveVW} presents an extension for
ASCL, a specification language for C programs. This ASCL extension introduces
a specification clause for relational properties, including \texttt{\textbackslash{}call}
operator, which is used to refer to several function invocations.
Both JML and Frama-C extension are not directly applicable for reactive systems,
and are not based on traces.

\section{Concrete and Generalised Test Table}
\label{sec:concrete-generalised}

As \rtts inherit their table-based syntax and their intuitive semantic from
\gtts~\cite{Beckert2017}, we outline the concept of \gtts in this section.

\begin{wrapfigure}[13]{R}{5cm}
  \vspace{-2em}
  \centering
  \begin{tabular}{c|rrr|rrr|c}
    & \multicolumn{3}{c}{Inputs} &
      \multicolumn{3}{|c|}{Outputs}& \\

      \# & $A$        & $B$              & $C$      & $X$       & $Y$           & $Z$           & \COLTIME \\
  \toprule
  0      & 1     & 1              & 2      & 0       & 0           & 5         & 1        \\
  1      & 0      & $3$ & $3$   & $6$  & 6        & 5 & $7$     \\
  2      & 1      & $4$         & 2    & 2 & 8 & 5    & 2      \\
\bottomrule
  \end{tabular}
  \caption{Example for a concrete test table with three input variables $A, B,
    C$ and three output variables $X, Y, Z$; it describes a test case of
    10~cycles.}
  \label{fig:concrete}
\end{wrapfigure}

\Gtts are derived from from concrete test tables, a table-based description for
test cases used in industry.
The table rows correspond to the consecutive steps of the test. The columns 
each correspond to an input or an output variable of the system under
test. In addition, there is an extra column \COLTIME{} whose entries denote
how often each row is to be repeated.
In a concrete table, the cells contain concrete values.
Hence, a concrete test table describes only one specific test case.

Fig.~\ref{fig:concrete} shows an example for a simple concrete test table. In
this example, all variables are of type integer; in general, other types, such
as Boolean variables, are also possible.

In contrast, a \gtt\ describes a family of concrete test tables by inserting
Boolean expression instead of concrete values within the tables cells.
The Boolean expression are built with the usual logical (${\land}, {\lor}$
etc.), arithmetical (${+},{*}$ etc.)\ and comparison (${<},{\geq}$ etc.)\
operators over the input and output variables.
Moreover,  references to values from previous cycles are possible; e.g.,
$X[-n]$~denotes the $X$'s value $n$~cycles earlier.
Also, expressions can contain global variables that have the same value
wherever they occur.

\setlength\intextsep{0pt}
\begin{wrapfigure}[12]{R}{7cm}
  \centering
  \begin{tabular}{ll}
    Abbrev. & Constraint                            \\
    \toprule
    $n$     & $X = n$                               \\
    $<n$    & $X<n$  (same for $>,\leq,\geq, \neq$) \\
    $[m,n]$ & $X \geq m\land X  \leq n$             \\
    $\alpha,\beta $  & $\alpha \land \beta$                           \\
    \DC{} & $X=X$  (don't care) \\
    \bottomrule
  \end{tabular}
  \caption{Constraint abbreviations ($X$~is the name of the variable that the
    cell corresponds to; $n,m$~are arbitrary expressions of type integer;
    $\alpha, \beta$ are abbreviations or formulae).}
  \label{fig:mnemonics}
\end{wrapfigure}
There are a number of abbreviations that we allow in expressions
for better readability and ease of use, see Fig.~\ref{fig:mnemonics}.
For example, the cell content 
$[n,m], {\neq}Z/2$ specifies that the value of~$X$ is
in the interval $[n,m]$ and not equal to the half of~$Z$.
We write ``\DC{}'' to allow an arbitrary value (``don't care'').

In a \gtt, the cells of the \COLTIME{}
column can also also contain constraints instead of concrete values. But,
here, we  only allow
intervals $[n,m]$ or $[n,\mbox{\DC{}}]$, where $n,m$ are concrete values,
and the special symbol~\strongrep{}.
An interval specifies the lower and upper bounds of row applications.
If the upper bounds is \DC{}, the number of row applications is arbitrary but
finite.
In contrast, \strongrep{} requires an infinite repetition.
It is possible that a test can continue to repeat a row or alternatively
progress to the next one if the constraints of both rows are satisfied. In such
cases, the user can enforce the test to progress to the next row by adding the
flag $p$ to the duration cell, e.g., writing $\dwait{[n,m]}$ instead of~$[n,m]$.

Multiple consecutive rows can be grouped to be repeated as a block.
Every group has its own additional \COLTIME{} constraint.
Hence, with row groups one can express repetitive patterns that span over more
than one row and may also include optional sub-patterns.
Row groups can be nested, i.e., a group may contain other groups and rows (but
they cannot partially overlap).

\begin{wrapfigure}[7]{R}[0cm]{7cm}
  \centering
  \scalebox{\scaleTables}{%
    \begin{tabular}{c|ccc|ccc|cp{6mm}p{6mm}}
      & \multicolumn{3}{c}{Inputs} & \multicolumn{3}{|c|}{Outputs} & \multicolumn{3}{c}{\duration}  \\

                                       \# & $A$      & $B$   & $C$    & $X$      & $Y$      & $Z$         &                                         \\
      \toprule
      0                                   & 1        & 1     & 2      & 0        & 0        & \DC         & 1          & {\rowgroupdurationd{0}{2em}{2}{\DONTCARE}} & \rowgroupdurationd{0}{3em}{3}{$\geq1$} \\
      1                                   & \DC      & $p$   & $p$    & $=\!2*p$ & $X$      & $Z[-1]$     & $\ge6$     &                            \\
      2                                   & \DC      & $p+1$ & \DC    & $[0,p]$  & $>Y[-1]$ & ~$2*Z > Y$~ & 1 &                            \\
    \end{tabular}}
  \caption{Example for a generalised test table with a global variable $p$.}
  \label{fig:gtt}
\end{wrapfigure}
Fig.~\ref{fig:gtt} shows an example of a simple generalised test table,
incorporating the generalisation concepts described above. Note that the
concrete table depicted in Fig.~\ref{fig:concrete} is one of the possible
instances of the generalised test table given in Fig.~\ref{fig:gtt}, achieved by
instantiating the global variable~$p$ with the value~$3$.

\section{Relational Test Tables: Syntax}
\label{sec:syntax}

In this section, we introduce the extensions that allow specifying
relational properties  and turn \gtts into \rtts{}.

We use a single \rtt to describe a class or family of relational test cases, 
which test for a relational property. There can be several \rtts specifying 
different scenarios, but each of them refers to all system traces. A \rtt has
one column for each input and each output variable of each of the
traces. There is a single duration column shared by all traces. 
Moreover, we add the following concepts:
\begin{inparaenum}[(a)]
\item relational references, i.e., references from one part of the table that
  corresponds to one system trace to parts of the table that correspond to
  some other trace, and
\item trace stuttering (pausing of trace), which allows to synchronise traces
  that do not proceed in lock-step.
\end{inparaenum}

\paragraph{Relational references}
We assume that names have been declared (outside the table) for the traces
that are to be put into relation. These can be, for example, \emph{old} and
\emph{new} in the case of regression verification (see the example in
Fig.~\ref{fig:rttexample}), or $p_1,\ldots,p_k$ in the more general case of
$k$~traces. 

A variable $X$ in a trace~$p$ is denoted by $\other[p]{X}$. As in \gtts, we
use the notation $\other[p]{X}[-n]$ to refer to the value of~$X$ in trace~$p$
in an earlier cycle ($n$~steps earlier).

As the case of two traces is very prominent,
we allow the abbreviation~``$\other{X}$'' (the trace identifier omitted) 
to refer to $X$ in the ``other'' trace if there are only two traces.
The notation ``\other{}'', where the variable name is also omitted, 
references the same variable in the other trace, i.e., 
\other{} equals \other[q]{X} if
it is used in the table column for variable \other[p]{X} and $p,q$ are the
only two traces.
Additionally, we keep the old notion: a simple name $X$ refers to the variable
$X$ from the same trace. See Table~\ref{fig:relmnemonics} for an overview of
notations for relational references.

\begin{wrapfigure}[9]{r}{7cm}
  \centering
  \begin{tabular}{cl}
    Notation                 & Value if used in column for $\other[p]{X}$                  \\
    \toprule
    $\other[q]{Y}$                     & $Y$ in trace $q$   \\
    $\other[\phantom{p}]{Y}$           & $Y$ in the other trace \\
    $\other[q]{\phantom{Y}}$           & $X$ in trace $p$ \\
    $\other[\phantom{p}]{\phantom{Y}}$ & $X$ in the other trace \\
    $Y$                                & $Y$ in trace $p$         \\
    \bottomrule
  \end{tabular}
  \caption{Notations for relational references.}
  \label{fig:relmnemonics}
\end{wrapfigure}

\paragraph{Trace pausing (stuttering)}
Synchronisation is an important issue in specifying and proving relation
program properties, in particular for regression
verification~\cite{DBLP:conf/kbse/FelsingGKRU14}.
If traces are asynchronous and do not run in lock-step, we need the
possibility to express in \rtts, where synchronisation points are and
which trace is supposed to wait for the other
trace(s), i.e., which trace should be paused (we do not distinguish between a
trace that is paused and a trace that is stuttering).
We enrich \rtts\ with a pause column for each trace.
Table cells in pause columns contain Boolean expressions. If that expression
evaluates to true in a cycle, the trace does not proceed and its input and
output values remain frozen.
For readability, we use the icon \stuttering{} for 
\texttt{TRUE} in pause columns, 
and leave a cell blank or  use \play{} for 
\texttt{FALSE} (non-stuttering).

\begin{figure}[t]
  \centering
  \resizebox{\textwidth}{!}{%
  \begin{tabular}{c|cc|ccc|cc|ccc}
    \# & \multicolumn{2}{c|}{Pause} %
       & \multicolumn{3}{c|}{Input} %
       & \multicolumn{2}{c|}{Output} %
       & \multicolumn{3}{c}{\duration}
       \\

       ~ & old                  & new & ~\other[new]{\mathit{WP}}~ & ~\other[old]{\mathit{WP}}~ & ~\other[new]{\mathit{Release}}~
         & ~\other[new]{\mathit{Press}}~ & ~\other[new]{\mathit{State}}~
 \\ \toprule

         \gttrow & \play       & \play & \other & \FALSE    & \DONTCARE & \other    & \other, =\v{Free}    & \dwait{-} &  & \rowgroupduration{6}{\strongrep} \\
         \gttrow & \play       & \play & \other & \TRUE     & \DONTCARE & \other    & \other,=\v{Stamping} & 1         &                                     \\
         \gttrow & \play       & \play & \other & \FALSE    & \DONTCARE & \other    & \other               & \DONTCARE &                                     \\
         \gttrow & \stuttering & \play & \other & \DONTCARE & \DONTCARE & \DONTCARE & \v{Error}            & \DONTCARE & \rowgroupduration{2}{[0,1]}         \\
         \gttrow & \stuttering & \play & \other & \DONTCARE & \TRUE     & \DONTCARE & \DONTCARE            & 1                                               \\
       \end{tabular}
     }
  \caption{Example for a \rtt}

  \label{fig:rttexample}
\end{figure}

\paragraph{Example}
We  illustrate our extension to \rtts with a small example. The goal is
to apply regression
verification to a small part of an automated production system, a stamping system for imprinting work pieces.
Stamping in the new version of the system 
should have the same behaviour as in the 
old version, except that there is an additional error handling routine.
If a work piece is inserted into the stamp (input variable~$\mathit{WP}$ gets
the value \texttt{TRUE}), the stamp
is pressed against the work piece (initiated by setting the output
$\mathit{Press}$ to \texttt{TRUE}) and the stamp signals when
it is ready ($\mathit{State}$).
The new revision is extended by a diagnostic sensor,
which recognises mis-stamping. If such an error is indicated,
the stamp needs to be  inspected and cleaned by an operator, who afterwards
releases the system from the error state.
Fig.~\ref{fig:rttexample} shows a \rtt capturing this behaviour. The normal
behaviour in the new system version is (only) described relationally, i.e., by
referring to the old version (rows~0--3). In addition, the table describes the
error handling behaviour without referring to the old version (rows~4--5).
In row~0, the table states that both systems should signal a free stamp
(variable~$\mathit{State}$) until
a work piece is inserted. The progress flag is set in the duration columns of
row~0 to ensure that the test case proceeds as soon as possible.
When a work piece present (row~1), the stamping process starts
(\v{State=Stamping}) for a unspecified amount of time (row~2).
Up until (and including) row~2 we expect that behave equally.
Then, when an error occurs, the equivalence requirement in row~2 fails as
the old revision is not aware of errors, and the test case proceeds to row~3.
The old revision is paused (\stuttering\ is set) until release by the operator
indicated by $\other[new]{\mathit{Release}}=\v{TRUE}$ (row~4).
The row group consisting of rows~3 and~4 makes the error handling optional.
The complete specification is repeated infinitely often.
Not all variables from the interface of both reactive systems have an own
column.
Some omitted variables are specified by indirectly, i.e. \other[old]{Press} and
\other[old]{State} via the columns of $\mathit{new}$.
\section{Relational Test Tables: Semantics}
\label{sec:semantics}

We define the semantics of \rtt, i.e., what it means for a system to conform
to a table, by reduction to the notion of conformance defined for
\gtt~\cite{Beckert2017}.

\paragraph{Conformance for \gtt}
A \gtt ${\cal T}$ describes a -- possibly infinite -- set of concrete test tables, which is obtained by
unwinding the rows and instantiating the cells with all sets of values that
satisfy the table's constraints.

Intuitively, a system trace~$t$  conforms to a table if (a)~it 
conforms to one of the concrete test
tables of~${\cal T}$ or (b)~none of the tests of~${\cal T}$ covers 
the input values of~$t$.
In~\cite{Beckert2017}, conformance  is formally defined by the outcome of
a two-party game of a challenger against the reactive system.
The challenger selects input stimuli according to the input
constraints, while the system's response must adhere to the output constraint.
A play of this game forms a concrete test table.
The first party violating a constraint looses the play.
The system also wins if
the current play is a (complete) concrete table test table of~${\cal T}$.

We distinguish two conformance levels: \emph{strict} and \emph{weak}.
A system \emph{strictly} conform to a \gtt~$\mathcal T$ iff it represents
a winning strategy, i.e., iff it wins against every possible challenger
strategy. 
The system \emph{weakly} conforms to a \gtt\ if its strategy never looses (but
possibly plays an infinite game without ever winning).

\paragraph{Conformance for \rtts}
For the reduction of conformance for \rtts to that for \gtts, we need to
\begin{inparaenum}[(a)]
\item handle stuttering,
\item combine several reactive programs into a single reactive product program, and
\item resolve relational references to other program runs (traces).
\end{inparaenum}

We handle stuttering by extending each of the reactive systems~$P_j$ by a new
implicit input variable~$stutt$.
If $stutt$ is true, the system ignores the input variables~$i$ and immediately
returns the same output as in the last cycle without changing its state.
The augmented reactive system $P'_j$ is defined as following
\begin{align}
  P'_j(stutt,i) &:= \texttt{if } \neg stutt \texttt{ then } P_j(i) \texttt{ else } \mathit{skip}  \texttt{ endif}\enspace.
\end{align}

We build the reactive product system $P'$ of the augmented reactive systems
$P'_1, \ldots, P'_n$ (cf.~\cite{Barthe2004}):
\[
  P'(i_1, \ldots, i_n) := (P_1(i), \ldots, P_n(i))\enspace,
\]
which is a parallel isolated execution of the reactive systems~$P_j'$.
The traces of $P'$ are the Cartesian products of the traces of the~$P'_j$.
The relational reference \other[p]{X} are rewritten to refer to the correspoding
variable in the product system.
We define the \rtt{} conformance for sequence of reactive systems
(cf.~\cite[Def.~5]{Beckert2017}):
\begin{definition}[Relational Conformance]
  \label{def:relconformance}
  A sequence of reactive systems $P_j: I^\omega \to O^\omega$
  ($1\leq j\leq n$) \emph{strictly conforms} to a \rtt~${\cal T}$ iff
  the product program $P'$ strictly conforms to the \gtt ${\cal T}'$, i.e., 
  is a winning strategy for
  the conformance game as defined in~\cite[Def.~5]{Beckert2017}. Analogously,
  it \emph{weakly conforms} to ${\cal T}$ iff the strategy never
  loses.
\end{definition}
The \gtt $T'$ is derived from the \rtt $T$ to the match the transformation
rules. In particular, the pause columns becomes an input columns, and the column
variables referring to the corresponding variables in the product reactive
system.

\section{Application Scenarios}
\label{sec:applicaton}

In this section, we show the applicability of \rtts using scenarios of the
Pick-and-Place Unit~(PPU) community demonstrator~\cite{Vogel-Heuser2014,BeckertUlbrichEtAl2015}.
The demonstrator consist of a magazine for providing new work pieces, a stamp for
imprint, a conveyor belt and a crane for transportation of 
work pieces.
All run times (wall clock) given below are a median of five samples where
conformance to the \rtt has been verified on an Intel i5-6500 3.20GHz with the
model checker nuXmv version 1.1.1~\cite{DBLP:conf/cav/CavadaCDGMMMRT14} and
IC3~\cite{4401997} for invariant checking.
All files of the verification are available at
 \url{https://formal.iti.kit.edu/rttreportarxiv}.

\subsection{Regression and Delta Verification}
\label{sec:rvdv}

\begin{figure}[t]
  \centering
  \includegraphics[width=\textwidth]{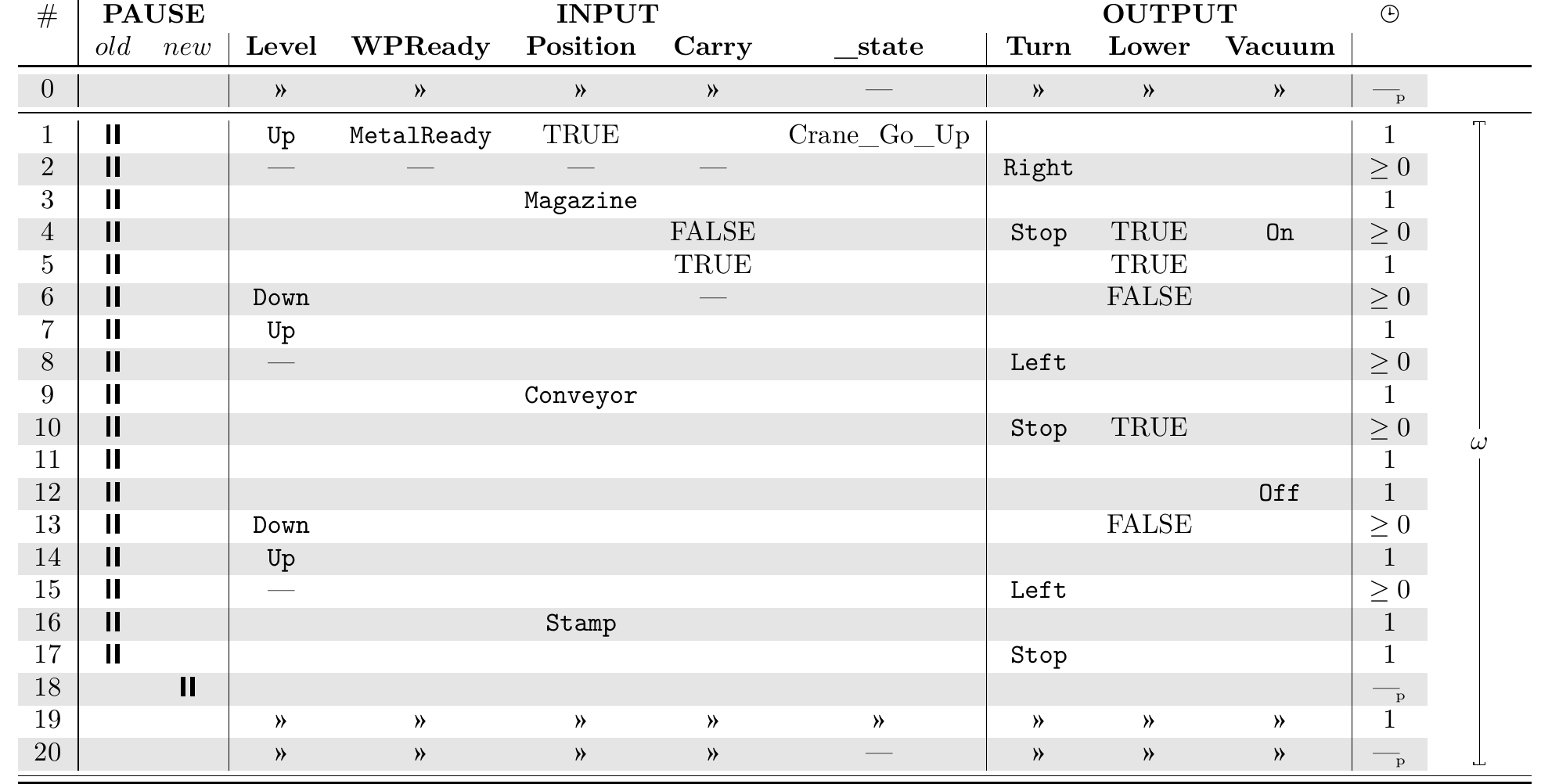}
  \caption{Combination of regression and delta verification. For presentation
    reason, we omitted trace reference in the column header. All columns
    belonging to the $new$ trace. }
  \label{fig:rvdv}
\end{figure}

\paragraph{Scenario}
In this scenario, we demonstrate a combination of regression verification and delta
verification~\cite{Ulewicz16} for two software revisions.
While regression verification proves the equivalence
for the common part of system 
behaviour, delta verification ensures functional correctness for the
differences.
This scenario is based on the evolution step from the third to the fifth software
revision of the PPU, which introduces an optimisation for work piece
throughput:
The new software revision makes use of 
the waiting time while a piece is stamped to deliver a new
work piece from the magazine to the conveyor belt.
The old revision waits for the stamp to finishing the imprint.

\paragraph{Table}
The \rtt  in Fig.~\ref{fig:rvdv} contains rows for the following 
input and output variables:
\begin{inparaenum}[(a)]
\item \emph{Level} indicating the position of the crane (\texttt{Up}, \texttt{Down},
  \texttt{Unknown}),
\item \emph{WPReady} signalling whether a \mbox{(non-)}metal 
work piece is ready at
  the magazine,
\item \emph{Position} of the crane (\texttt{Magazine}, \texttt{Stamp}, etc.),
\item \emph{Carry} signalling whether a work piece has been picked up,
\item the current \emph{\_state} of the internal state machine,
\item \emph{Turn} determining the move direction of the crane (\texttt{Stop}, \texttt{Left}, \texttt{Right}),
\item the desired position of the suction cup (\emph{Lower}), and
\item whether the suction cup should hold a work piece.
\end{inparaenum}
The table specifies that the $\mathit{old}$ (third) revision and the
$\mathit{new}$ (fifth) revision behave
equally (Row~0, Row~19, Row~20), except for the phase in which the optimisation
is occurs.
During the optimisation phase, the trace of the $\mathit{old}$ revision stutters (Row~1
to~17) while the $\mathit{new}$ trace moves the crane to the magazine, picks up the work
piece, delivers it to the conveyor belt, and moves the crane back to the stamp.
This sequence is described as a functional specification.
In Row~18, we pause the $\mathit{new}$ trace and let the $\mathit{old}$ trace run until both
traces are synchronised again on the same internal state $\_state$.
This is required, because the waiting duration for imprinting is hard-coded and
cannot be changed.

\paragraph{Verification}
We proved system conformance to the \rtt for the 
function block of Crane module, resulting
into a product program with 726~lines of code and 72~variables.
Verification of weak conformance took 5.66~seconds with a state size of
209~bits (software and table).
For the verification we decreased the waiting durations of the timers.

\subsection{Information Flow}
\label{sec:information-flow}

\begin{figure}[t]
  \centering
  \includegraphics[width=\textwidth]{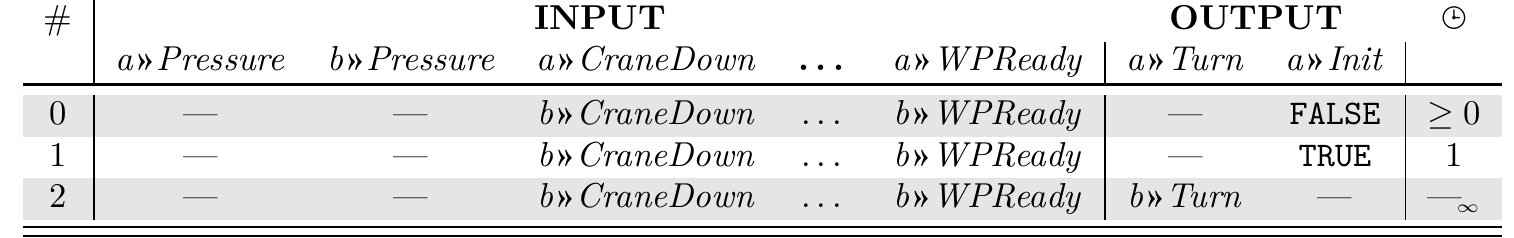}
  \caption{Information-flow property:  the input $\mathit{Pressure}$ should not have an
    influence on the output $\mathit{Turn}$.}
  \label{fig:if}
\end{figure}

\paragraph{Scenario and Table}
In this scenario, we verify that there is no information flow from the suction
pressure ($\mathit{Pressure}$) to the crane movement ($\mathit{Turn}$).
More precisely, the table in Fig.~\ref{fig:if} describes that the non-interference is only required after
the initialisation of the system (Row~2, $\mathit{Init}$).
In all rows, we enforce that all input variables besides 
$\mathit{Pressure}$ are equal in both runs.
Row~2 expresses the non-interference property: For any two runs with arbitrary
values for $\mathit{Pressure}$, the output
$\mathit{Turn}$ is the same.
Therefore, $\mathit{Turn}$ is only determined by the other input variables.

Unfortunately, a monitor function block stops the crane if the suction pressure
is outside the expected range, i.e., if $\mathit{Pressure} \not\in [1,9999]$.
Therefore, the software does in fact not conform to the \rtt in Fig.~\ref{fig:if}.
This unintended outcome can be fixed by limit the (unknown) values of $\mathit{Pressure}$ in both traces to the range
$[1,9999]$.

\paragraph{Verification}
We used the complete fifth revision of the PPU, including the function blocks
for all components.
The product program has 1758~lines of code with 266~variables.
For the interfering version, the model checker needs 25.93~seconds to find a counter example.
Proving conformance w.r.t.\ the 
fixed -- non-interfering -- specification takes 122.26~seconds.
The size of the state space is 373~bits.

\section{Conclusion}
\label{sec:conclusion}
In this paper, we have presented \rtts, 
a canonical extension of the \gtt concept for the
specification of relational properties for reactive systems.
The semantics of \rtt\ is defined by reduction to that of \gtts, whereas the syntax
is a genuine extension to \gtts.
For future work, the next step is the adaption of time synchronisation points to
increase the flexibility between traces. For example, allowing the specification
of parallel runs with all possible interleavings.

\bibliography{references}

\begin{thebibliography}{10}
\providecommand{\url}[1]{\texttt{#1}}
\providecommand{\urlprefix}{URL }

\bibitem{Barthe2004}
Barthe, G., D'Argenio, P.R., Rezk, T.: Secure information flow by
  self-composition. In: Proceedings. 17th IEEE Computer Security Foundations
  Workshop, 2004. pp. 100--114 (June 2004)

\bibitem{DBLP:journals/corr/abs-1906-09899}
Barthe, G., Eilers, R., Georgiou, P., Gleiss, B., Kov{\'{a}}cs, L., Maffei, M.:
  Verifying relational properties using trace logic. CoRR  abs/1906.09899
  (2019), \url{http://arxiv.org/abs/1906.09899}

\bibitem{Beckert2017}
Beckert, B., Cha, S., Ulbrich, M., Vogel-Heuser, B., Weigl, A.: Generalised
  test tables: A practical specification language for reactive systems. In:
  Polikarpova, N., Schneider, S. (eds.) Integrated Formal Methods. pp.
  129--144. Springer International Publishing, Cham (2017)

\bibitem{BeckertUlbrichEtAl2015}
Beckert, B., Ulbrich, M., Vogel-Heuser, B., Weigl, A.: Regression verification
  for programmable logic controller software. In: 17th International Conference
  on Formal Engineering Methods (ICFEM 2015). LNCS, vol. 9407, pp. 234--251.
  Springer (Dec 2015)

\bibitem{Blatter2016DeductiveVW}
Blatter, L., Kosmatov, N., Gall, P.L., Prevosto, V.: Deductive verification
  with relational properties. ArXiv  abs/1606.00678 (2016)

\bibitem{4401997}
Bradley, A.R., Manna, Z.: Checking safety by inductive generalization of
  counterexamples to induction. In: Formal Methods in Computer Aided Design,
  2007. FMCAD '07. pp. 173--180 (Nov 2007)

\bibitem{DBLP:conf/cav/CavadaCDGMMMRT14}
Cavada, R., Cimatti, A., Dorigatti, M., Griggio, A., Mariotti, A., Micheli, A.,
  Mover, S., Roveri, M., Tonetta, S.: The {nuXmv} symbolic model checker. In:
  Computer Aided Verification (CAV). pp. 334--342. LNCS 8559, Springer (2014)

\bibitem{DBLP:conf/post/ClarksonFKMRS14}
Clarkson, M.R., Finkbeiner, B., Koleini, M., Micinski, K.K., Rabe, M.N.,
  S{\'{a}}nchez, C.: Temporal logics for hyperproperties. In: Abadi, M.,
  Kremer, S. (eds.) Principles of Security and Trust - Third International
  Conference, {POST} 2014, Held as Part of the European Joint Conferences on
  Theory and Practice of Software, {ETAPS} 2014, Grenoble, France, April 5-13,
  2014, Proceedings. Lecture Notes in Computer Science, vol. 8414, pp.
  265--284. Springer (2014), \url{https://doi.org/10.1007/978-3-642-54792-8_15}

\bibitem{Clarkson2010}
Clarkson, M.R., Schneider, F.B.: {Hyperproperties}. Journal of Computer
  Security  18(6),  1157--1210 (2010)

\bibitem{Denning}
Denning, D.E.: A lattice model of secure information flow. Commun. ACM  19(5),
  236--243 (May 1976), \url{http://doi.acm.org/10.1145/360051.360056}

\bibitem{DBLP:conf/kbse/FelsingGKRU14}
Felsing, D., Grebing, S., Klebanov, V., R{\"{u}}mmer, P., Ulbrich, M.:
  Automating regression verification. In: Crnkovic, I., Chechik, M.,
  Gr{\"{u}}nbacher, P. (eds.) {ACM/IEEE} International Conference on Automated
  Software Engineering, {ASE} '14, Vasteras, Sweden - September 15 - 19, 2014.
  pp. 349--360. {ACM} (2014), \url{http://doi.acm.org/10.1145/2642937.2642987}

\bibitem{PaVy:16}
Pakonen, A., Pang, C., Buzhinsky, I., Vyatkin, V.: User-friendly formal
  specification languages – conclusions drawn from industrial experience on
  model checking. In: IEEE International Conference on Emerging Technologies
  and Factory Automation (ETFA 2016). Berlin, Germany (2016)

\bibitem{DBLP:conf/fm/SchebenS14}
Scheben, C., Schmitt, P.H.: Efficient self-composition for weakest precondition
  calculi. In: Jones, C.B., Pihlajasaari, P., Sun, J. (eds.) {FM} 2014: Formal
  Methods - 19th International Symposium, Singapore, May 12-16, 2014.
  Proceedings. Lecture Notes in Computer Science, vol. 8442, pp. 579--594.
  Springer (2014), \url{https://doi.org/10.1007/978-3-319-06410-9_39}

\bibitem{Ulewicz16}
Ulewicz, S., Ulbrich, M., Weigl, A., Kirsten, M., Wiebe, F., Beckert, B.,
  Vogel-Heuser, B.: A verification-supported evolution approach to assist
  software application engineers in industrial factory automation. In: IEEE
  International Symposium on Assembly and Manufacturing (ISAM). pp. 19--25.
  Fort Worth, USA (2016)

\bibitem{Vogel-Heuser2014}
Vogel-Heuser, B., Legat, C., Folmer, J., Feldmann, S.: {Researching Evolution
  in Industrial Plant Automation: Scenarios and Documentation of the Pick and
  Place Unit}. Tech. Rep. TUM-AIS-TR-01-14-02 (2014),
  \url{https://mediatum.ub.tum.de/doc/1208973/1208973.pdf}

\bibitem{DBLP:journals/tosem/YiQTR15}
Yi, J., Qi, D., Tan, S.H., Roychoudhury, A.: Software change contracts. {ACM}
  Trans. Softw. Eng. Methodol.  24(3),  18:1--18:43 (2015),
  \url{http://doi.acm.org/10.1145/2729973}

\end{thebibliography}
\end{document}